%% file: paper.tex
\documentclass[conference]{IEEEtran}
\IEEEoverridecommandlockouts

\usepackage{tikz}
\usepackage{textcomp}
\usepackage{lipsum}

\usepackage{cite}
\usepackage{float}
\usepackage{dblfloatfix}
\usepackage{amsmath,amssymb,amsfonts}
\usepackage{algpseudocode}
\usepackage{algorithm}
\usepackage{array,multirow,graphicx}
\usepackage{graphicx}
\usepackage{textcomp}
\usepackage{xcolor}
\def\BibTeX{{\rm B\kern-.05em{\sc i\kern-.025em b}\kern-.08em
    T\kern-.1667em\lower.7ex\hbox{E}\kern-.125emX}}
\usepackage{subfig}
\usepackage{booktabs}
\usepackage{enumitem}
\usepackage{threeparttable}
\usepackage[multiple]{footmisc}
\usepackage{balance}
\usepackage{xcolor,colortbl}
\usepackage[hyperfootnotes=false]{hyperref}

\definecolor{Gray}{gray}{0.85}
\newcolumntype{a}{>{\columncolor{Gray}}c}

\DeclareMathAlphabet{\altmathcal}{OMS}{cmsy}{m}{n}

\begin{document}

\title{Khaos: Dynamically Optimizing Checkpointing for Dependable Distributed Stream Processing}

\author{
\IEEEauthorblockN{Morgan K. Geldenhuys\IEEEauthorrefmark{1},
Benjamin J. J. Pfister\IEEEauthorrefmark{1},
Dominik Scheinert\IEEEauthorrefmark{1},
Lauritz Thamsen\IEEEauthorrefmark{3}, and
Odej Kao\IEEEauthorrefmark{1}
}
\IEEEauthorblockA{Technische Universit{\"a}t Berlin, Germany, \{firstname.lastname\}@tu-berlin.de}
\IEEEauthorblockA{
\IEEEauthorrefmark{3}
University of Glasgow, United Kingdom, lauritz.thamsen@glasgow.ac.uk}
}

\maketitle

\makeatletter

\begin{abstract}
Distributed Stream Processing systems are becoming an increasingly essential part of Big Data processing platforms as users grow ever more reliant on their ability to provide fast access to new results. 
As such, making timely decisions based on these results is dependent on a system's ability to tolerate failure.
Typically, these systems achieve fault tolerance and the ability to recover automatically from partial failures by implementing checkpoint and rollback recovery. 
However, owing to the statistical probability of partial failures occurring in these distributed environments and the variability of workloads upon which jobs are expected to operate, static configurations will often not meet Quality of Service constraints with low overhead.

In this paper we present Khaos, a new approach which utilizes the parallel processing capabilities of cloud orchestration technologies for the automatic runtime optimization of fault tolerance configurations in Distributed Stream Processing jobs.
Our approach employs three subsequent phases which borrows from the principles of Chaos Engineering: establish the steady-state processing conditions, conduct experiments to better understand how the system performs under failure, and use this knowledge to continuously minimize Quality of Service violations. 
We implemented Khaos prototypically together with Apache Flink and demonstrate its usefulness experimentally.

\end{abstract}


\input{sections/01_introduction}
\input{sections/02_related_work}

\input{sections/03_approach}
\input{sections/04_evaluation}
\input{sections/06_conclusion}

\section*{Acknowledgment}

This work has been supported through grants by the German Ministry for Education and Research (BMBF) as BIFOLD (funding mark 01IS18025A) and WaterGridSense 4.0 (funding mark 02WIK1475D).

\bibliographystyle{IEEEtran}
\bibliography{paper}

\end{document}

%% file: sections/01_introduction.tex
\section{Introduction}

With the growing necessity to quickly process large volumes of unbounded data, Distributed Stream Processing (DSP) systems are becoming an increasingly essential part of data processing environments. 
It is here where events must traverse a graph of streaming operators to allow for the extraction of results, which are at their most valuable closest to the time of data arrival. 
Data streams are continuously generated in a variety of contexts such as sensors in IoT network, network monitoring, financial fraud detection, click stream analytics, spam filtering, and social media~\cite{IAM+19, NNG19}. 
DSP systems are not only required to offer near to real-time processing latencies at high throughput rates, but also to recover from the various types of failures that inevitably occur in these environments.

DSP jobs are, in principle, required to operate indefinitely on unbounded streams of continuous data and exhibit heterogeneous modes of failure as they continue to run over long periods of time~\cite{LWJ+17}. Consequently, DSP systems such as Apache Storm~\cite{TTS+14}, Apache Spark~\cite{ZMC+10}, or Apache Flink~\cite{CKE+15} feature high availability modes and fault tolerance mechanisms that allow for results to be consistent in the presence of partial failures. 
However, the complexity with which these systems are composed makes estimating how a system will perform through manual manipulation of the configuration sets a hard problem to solve. 
This is complicated by the fact that DSP jobs operating in an environment where streaming workloads change over time will make any static configuration selections obsolete in short order. 
This is especially relevant when considering critical jobs where the presence of Quality of Service (QoS) constraints dictate the minimum level of performance that is to be expected. 
It is therefore essential when optimizing the fault tolerance mechanism of DSP jobs to not only understand how configuration impacts upon recovery times as well as end-to-end processing latencies, but to ensure that the system is capable of reacting to changing workloads.

Checkpointing mechanisms are among the most popular and effective techniques for achieving fault tolerance in real world processing systems and consequently a number of methods for auto-configuration of checkpointing have been proposed.
Most of them try to optimize the checkpoint interval by means of predicting or utilizing failure rates~\cite{JHK20}, the Mean Time To Failure (MTTF)~\cite{Y74,D03,D06}, or recovery times~\cite{DBLP:conf/bigdataconf/GeldenhuysTK20}, whereas some approaches even employ advanced multi-level checkpointing~\cite{GKA+10,GMC+10,LPW+14,MBM+10,BTK+11,KMI+12}. 
Yet, the majority of such methods either assume static workloads, consider solely offline optimization, or are primarily designed for high-performance computing (HPC) environments, which renders them not suitable for real-world DSP systems.
Therefore, a workload-adaptive DSP configuration optimization approach, paired with a systematically attempt to evaluating DSP failure recovery performance executing in production-like environments, is needed.

In this paper we present Khaos, a novel approach for the automatic runtime optimization of DSP fault tolerance configurations. 
Borrowing from the principles of Chaos Engineering\cite{BBR+16}, Khaos achieves this by executing a three phase plan: Firstly, establishing the steady-state by recording and then analyzing the streaming workload of a targeted DSP job to identify interesting points for failure injection;
Secondly, by taking advantage of container orchestration cloud technologies to replicate multiple pipelines in parallel where the workload is replayed, and then utilizing fine-grained failure injection together with an anomaly detector trained on normal pipeline executions to measure recovery times across a range of configuration settings and throughput rates; and thirdly, by taking the information gathered during profiling to train analytical models to monitor for when recovery times and latencies would exceed user-defined QoS constraints and automatically optimize for better fault tolerance configurations at runtime. 
To determine if violations should be acted upon immediately or deferred for later, Khaos makes use of Time Series Forecasting (TSF) to predict future workloads.
Khaos is applicable for users willing to initially accept an increased level of resource utilization to ensure optimized execution over the longer term.

The remainder of the paper is structured as follows: 
Section II explores the related work regarding DSP systems and their fault tolerance mechanisms as well as adaptive checkpointing schemes.
Section III presents our approach to automatically optimizing the fault tolerance mechanism of targeted DSP jobs operating on variable workloads; Section IV describes our evaluation through performing two experiments; and Section V summarizes our findings.

%% file: sections/02_related_work.tex
\section{Related Work}

In this section we examine how fault tolerance is handled in three popular DSP systems as well as work most related to our own, aimed at adaptive checkpointing and failure injection. 

\subsection{Distributed Stream Processing Systems}

In DSP, checkpointing is the most popular fault tolerance mechanism for real world systems. One of the first widely used large-scale DSP systems is Apache Storm \cite{TTS+14}, which guarantees that in the event of failure, messages are re-processed by using a mechanism of upstream operator backup and message acknowledgements. Although such a mechanism does guarantee messages are processed at least once, it also results in duplicate messages passing through the system and, therefore, falls short of the exactly-once processing guarantees needed by many modern DSP jobs.

Apache Spark Streaming \cite{ZMC+10}, on the other hand, is designed around the idea of micro-batching where messages are grouped together in an attempt to overcome the overhead caused by message-level synchronization. Here, micro-batches either succeed or are recomputed when failures occur. It uses periodic checkpointing to truncate the RDD lineage graph and save both metadata and data to reliable storage. This technique allows for exactly-once processing of messages.

Apache Flink \cite{CKE+15} is a well-known DSP system, which likewise provides exactly-once processing guarantees through periodically creating and saving a distributed snapshot of the global state \cite{CFE+15}. It achieves this by passing streaming barriers through the execution graph from source to sink operators. At the arrival of a barrier, each operator performs a checkpoint of the local state and then passes the barrier on to all output edges. A checkpoint is considered complete when all operators have completed their individual checkpoints.

\subsection{Adaptive Checkpointing}

A number of approaches have been proposed that optimize the fault tolerance configuration parameters by finding an optimal checkpoint interval (CI) to improve performance. 
Our previous work~\cite{DBLP:conf/bigdataconf/GeldenhuysTK20} focused on predicting recovery times and then optimizing the CI with regards to a single user-defined QoS constraint. 
However, this approach is aimed at scenarios where jobs process a static workload, i.e. throughput does not change over time. 
In the closely related area of research, we published an approach which uses \textit{times series forecasting} to optimize the resource utilization of DSP jobs executing in environments where the workload is expected to change over time~\cite{Geldenhuys2022PhoebeQD}.
A group of approaches focuses on determining the mean time to failure (MTTF) of cluster nodes and then adaptively fitting a CI that minimizes the time lost due to failure \cite{Y74,D03,D06}. 
These approaches, however, are more appropriate to high-performance computing (HPC) clusters and batch processing jobs as they rely on jobs having a finite execution time as part of their calculations. 
Specific to DSP systems, \cite{JHK20} incorporates failure rates in an attempt to fit the CI based on the MTTF. 
However, unlike Khaos, optimizations are not performed at runtime and therefore dynamic workloads are not taken into account. 
Additionally, it does not incorporate the total time needed to recover to processing events at the latest timestamp nor consider any user-defined QoS constraints for its optimization step.

Other approaches have been proposed using multi-level checkpointing schemes to resolve the issue of checkpoint/recovery overhead \cite{GKA+10,GMC+10,LPW+14,MBM+10,BTK+11,KMI+12}. 
Different checkpointing levels are used, which in turn are more flexible than traditional single-level schemes as it can consider multiple-failure types with each having a different checkpoint/recovery cost associated. 
Likewise, differing checkpoint levels can be associated with different storage types, which is usually not possible with single-level checkpoints. 
In~\cite{DRV+17}, a two-level checkpointing model is proposed: checkpoint level 1 deals with errors with low checkpoint/recovery overheads such as transient memory errors, and checkpoint level 2 deals with hardware crashes such as node failures. 
These approaches are specific to HPC environments and need adaption before vendors can consider implementing them in DSP systems.

\subsection{Failure injection}
Here we present approaches which use failure injection as a means of gaining a greater insight into how systems perform when things go wrong.
Failure injection for distributed systems is realized in~\cite{HLP+09} using virtualization at a low level. 
Both machines and networks are virtualized, and a failure injection is performed by uncontrolled shutdown of machines. The authors whole approach is accurately described as an emulation platform. 
In contrast, Khaos utilizes modern container orchestration technologies for deploying DSP pipelines and thus eases the emulation furthermore.  
In another work, fault injection is used in~\cite{SGA+11} to assess the effectiveness of partial fault tolerance techniques in DSP applications. 
The authors identify four metrics that can be used to evaluate the impact of faults in different stream operators with respect to predictability and availability. 
To reduce the number of required fault injection targets when evaluating a target application, their framework pre-analyzes the data flow graph.
An approach that combines fault injection and data analytics is motivated in~\cite{PWT+17}. 
Here, a database of failures is populated during a profiling phase, and queried during execution of production DSP jobs. 
The database is specifically used to help identify root causes of failures observed by providing injected faults that generated similar processing flows.

Netflix injects failures into its production system using its Chaos Automation Platform (ChAP) and Failure Injection Testing (FIT) tools \cite{BHJ+19}.
ChAP performs Chaos experiments on a small, randomly selected percentage of live traffic. 
ChAP deploys two scaled-down clusters of the Netflix service, one to serve as the performance baseline, the other to serve as a canary group. 
FIT injects either error responses or increases latency to simulate failure scenarios \cite{BBH+16} for the canary group and the results are compared against the baseline. 
Results of the Chaos experiment are evaluated by performing anomaly comparisons to detect statistically significant deviations between the normal state and Chaos experiment \cite{BHJ+19}.

%% file: sections/03_approach.tex
\section{Approach}
This section describes in detail the various phases of our approach Khaos, right after a general overview.

\subsection{Idea Overview}

The configuration of a fault tolerance mechanism has a direct impact both on the performance and availability of any running DSP job. 
Khaos borrows from the principles of Chaos Engineering and takes advantage of the massively parallelizable capabilities of container orchestration cloud technologies to provide an automated runtime approach for tuning the fault tolerance configuration of targeted DSP jobs. 
It achieves this by first conducting parallel profiling runs where failures are injected into short-lived profiling jobs, each testing a variation of the configurations, and gathering metrics related to the latencies and recovery times. 
These results are then used to train two multivariate runtime models that, when combined, provide a mechanism whereby user-defined performance and availability constraints are monitored for violations.
Should this occur, the system can be automatically reconfigured to provide the best trade-off between constraints. 

The cost of reconfiguration should likewise be taken into account, as it requires a full restart of the job with several current DSP implementations, albeit without having to reprocess any messages that might have accumulated during the downtime.
This is because controlled restarts perform a system save immediately before making the change and therefore no consumer lag can build up.
In order to do this we make use of TSF to determine if a reconfiguration should be performed at the current point in time, or if the decision can be deferred to the next optimization cycle.
The logic behind this is that if the workload is expected to decrease substantially, then a reconfiguration is not necessary.
Overall, such an optimization approach is imperative when operating in an environment where the workload changes dynamically over time and any static configurations are essentially made obsolete immediately. 
In order to achieve this, our approach is subdivided into three distinct phases that are executed sequentially. 
Next we describe each of these phases and provide formal definitions.

\begin{figure*}
\centering
\subfloat[Phase 1: Establishing the Steady State.]{
  \includegraphics[width=56mm]{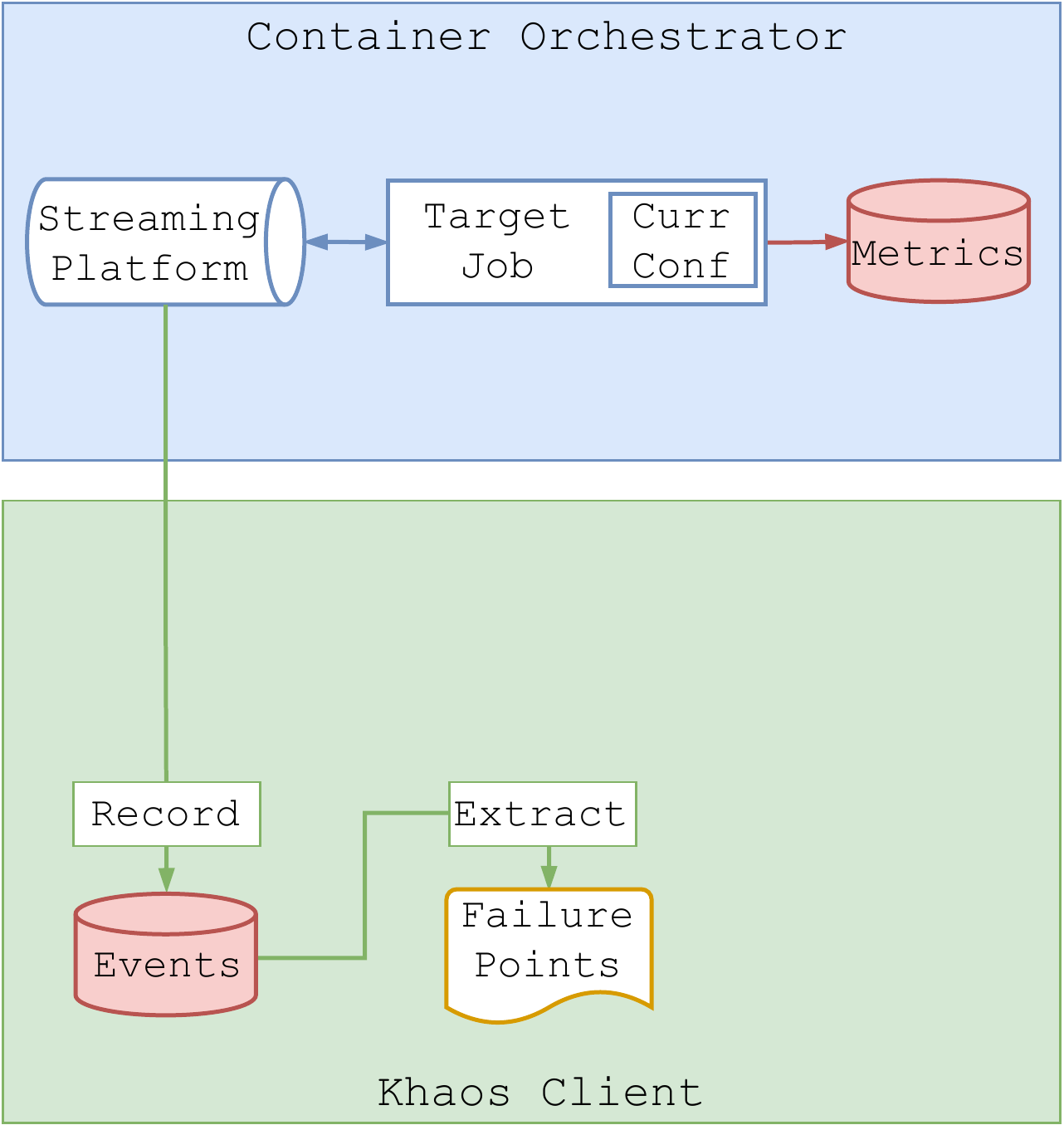}
}
\subfloat[Phase 2: Experimentation and Profiling.]{
  \includegraphics[width=56mm]{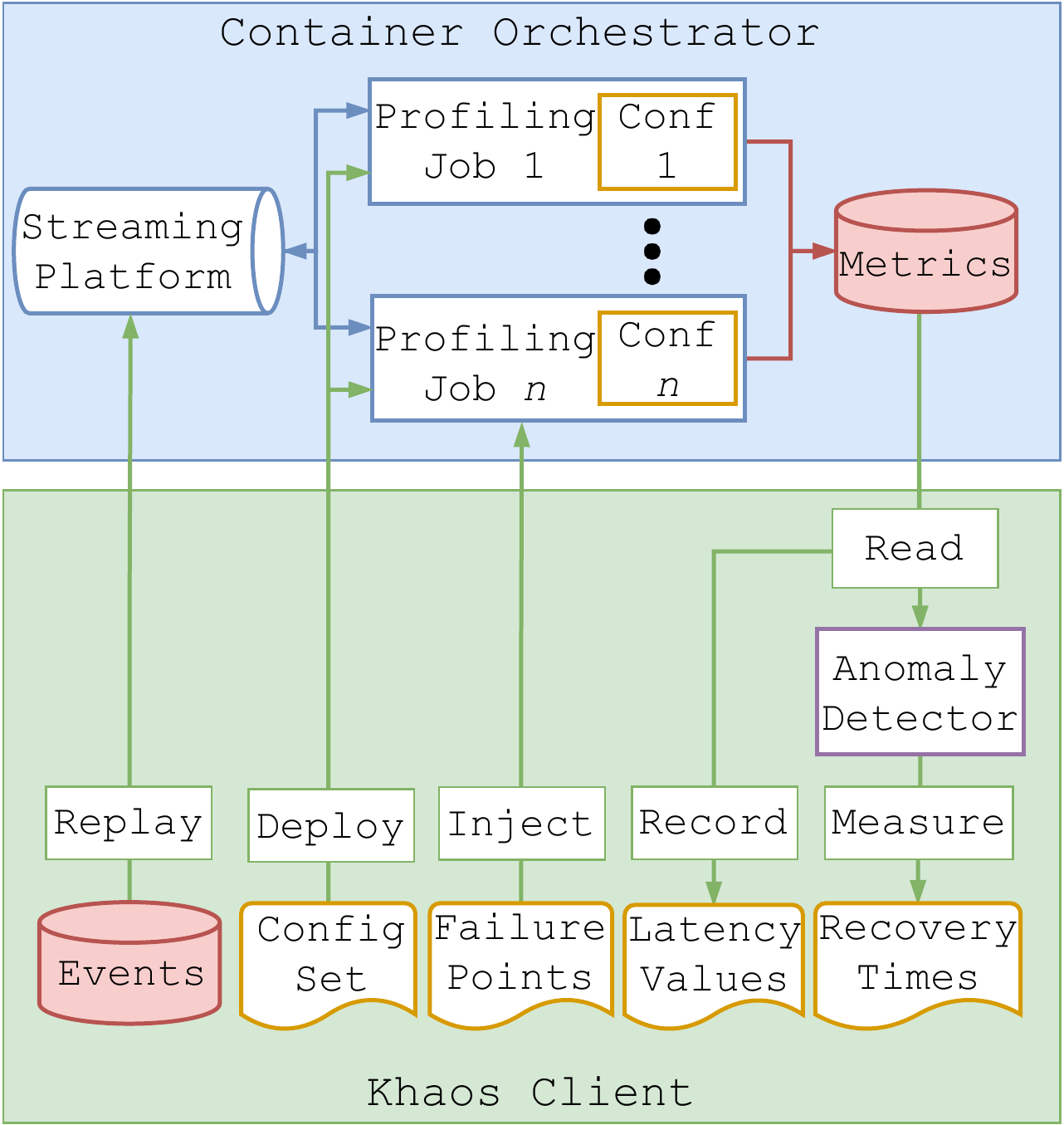}
}
\subfloat[Phase 3: Modeling and Runtime Optimization.]{
  \includegraphics[width=56mm]{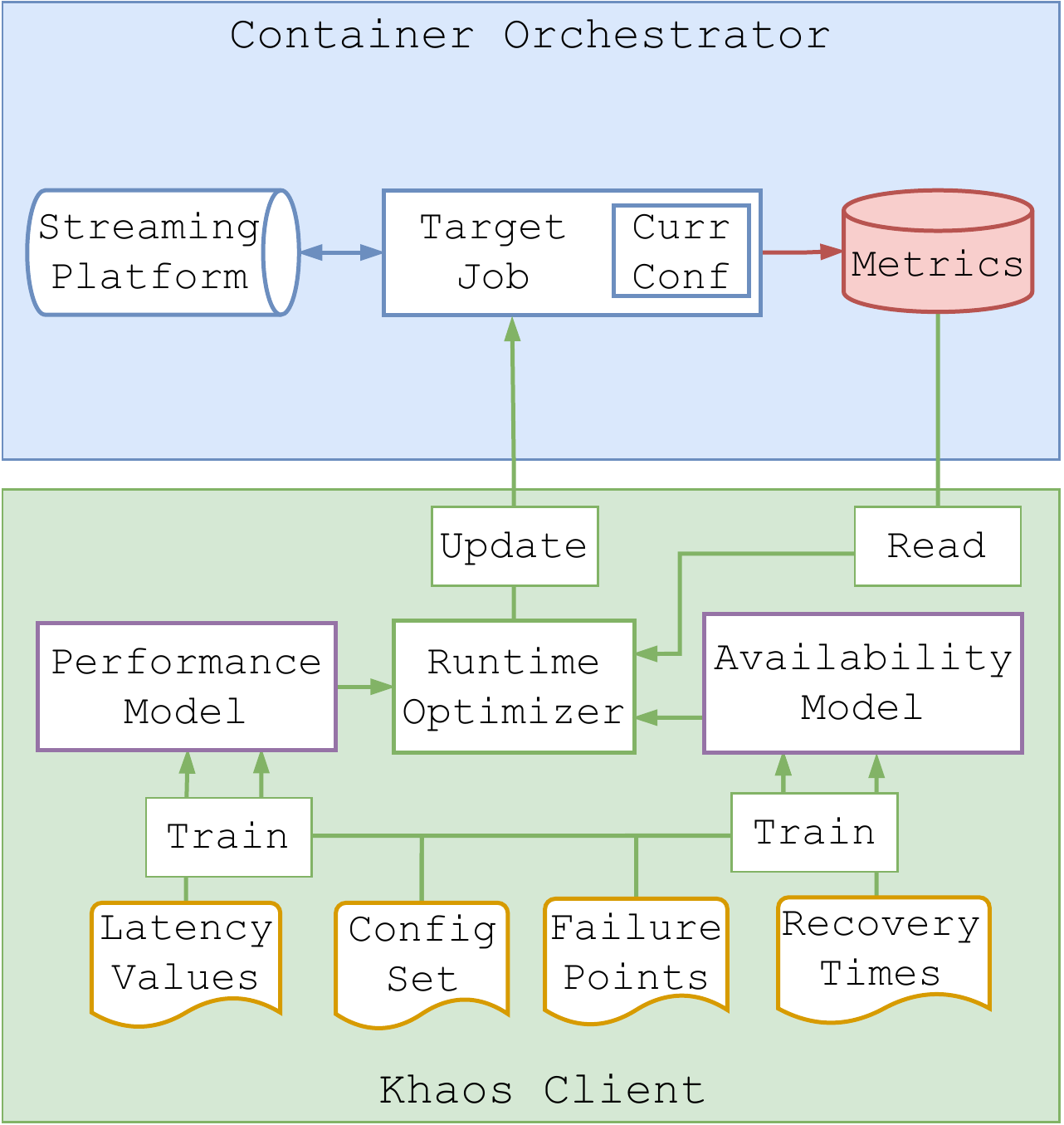}
}
\caption{Overview of Khaos.}
\label{fig:overview}
\end{figure*}

\subsection{Phase 1: Establishing the Steady State}

The first phase focuses on capturing and establishing the steady state that describes job performance under failure-free conditions. 
A graphical overview of this phase can be seen in \autoref{fig:overview}(a).
Consider a production-level DSP job a user would like to have adaptively optimized. 
The job is executing in a containerized environment, consuming a variable workload from a streaming platform, and metrics are gathered in a time series database. 
On initialization, Khaos connects to the streaming platform and begins recording the incoming event stream. 
It does this for a finite amount of time $k$ as defined by the user and ideally should contain the maximum range of throughput rates as expected under normal processing loads across that period.
For best results, DSP jobs processing a stationary data stream would best fit our approach.
When considering any single timestamp $t_i$ within this recording period, we can define the set of arriving events as $E^{(t_i)} = \{e_1^{(t_i)}, e_2^{(t_i)},\ldots, e_{n-1}^{(t_i)}, e_n^{(t_i)}\}$, where $n$ denotes the number of arriving events in that timestamp. 
Consequently, the full set $D$ of events arriving across all timestamps during the recording period is defined as

\begin{equation}
    D = \{E^{(t_1)}, E^{(t_2)},\ldots, E^{(t_{k - 1})}, E^{(t_k)}\}.
\end{equation}
 
As Khaos is concerned with injecting failures into a running system to gain an understanding of how recovery times differ across various configurations, it needs to select a set of points over the full range of observed throughput rates where failures can be injected when the dataset is replayed. 
Therefore, a continuous function $W$ of time $t$ is extracted from $D$ representing the workload over time and is defined by

\begin{equation}
  W(t) = |E^{(t)}| 
\end{equation}

This function is analyzed to find a set of equidistantly spaced throughput rates between the minimum and maximum observed workloads and their corresponding timestamp values. Importantly an averaging window is used to smooth the workload function and remove outliers. Formally, we identify 

\begin{equation}
    \begin{gathered}
        t_{\text{min}} = \text{arg min}_{0 \leq j \leq k} W(j)\\
        t_{\text{max}} = \text{arg max}_{0 \leq j \leq k} W(j)
    \end{gathered}
\end{equation}

as the points in time with minimum and maximum workload. Subsequently, we find $m$ equidistant points and arrive at a set $F$ of timestamps representing the \emph{failure points}, i.e.

\begin{equation}
    \begin{gathered}
        F = \{t_\text{min}, t_{\text{min} + h},\ldots, t_{\text{max} - h}, t_{\text{max}}\}, \\
        h = (t_{\text{max}} - t_{\text{min}}) / (m-1).
    \end{gathered}
\end{equation}

Since $F$ is a set of timestamps, the corresponding set of throughput rates $TR$ can be defined as
\begin{equation}
    TR = \{W(f)| f\in F\}.
\end{equation}

On the conclusion of phase 1, the steady state of the production DSP job has been explored and the dataset $D$, failure points $F$ and corresponding throughput rates $TR$ are available for use in the next phase. 

\subsection{Phase 2: Experimentation and Profiling}

The second phase focuses on performing experiments in an environment that is as similar as possible to production with the goal of gathering metrics data using the same DSP job but executed across a closed range of configuration settings. 
In order to do so we take advantage of container orchestration cloud technologies to rapidly replicate multiple profiling pipelines in parallel, record their operational latencies under differing throughput rates, and then use fine-grained failure injection together with an anomaly detector $A$ to measure recovery times. 
\autoref{fig:overview}(b) provides a graphical overview of this phase and illustrates the interactions between components.

Concerning configuration of the fault tolerance mechanism, we are most concerned with the checkpoint interval, defined as the frequency with which the checkpoint process is initiated. 
It is by varying the CI that we are able to configure the fault tolerance mechanism, where higher values represent possible longer recovery times and, conversely, lower values represent faster recovery times. 
Varying the checkpoint interval will have an impact on both the overall performance and availability of the system. 
For the CI, we seek to investigate a range of $z$ equidistantly spaced values given a minimum value and a maximum value. 
We effectively obtain a set of concrete configurations $C$ with $z = |C|$ analogue to the procedure for $F$. 
Thus, we can test all configurations in $C$ at the same time, i.e. arriving at as many deployments as configuration values. 

After the parallel profiling jobs have been instantiated, each timestamp in the failure points $F$ is registered with a built-in failure injector. 
This is so that when the timestamp, and therefore the associated throughput rate, is realized, Khaos will inject a failure concurrently into each one of the parallel deployments.
Once this has been completed, Khaos will begin replaying the dataset $D$ at the same rate at which it was recorded. All parallel profiling jobs will read from the same source to which the data is being replayed. It is important to note that this is a separate messaging queue to the one being consumed by the production job. 
At the point before injecting the failure, an average latency measurement is taken for each of the parallel deployments forming the set $L$ with 

\begin{equation}
    L = \{l_i^{(j)}| 1 \leq i \leq m, 1 \leq j \leq z\},
\end{equation}

where $l_i^{(j)}$ is the average latency measurement corresponding to the $i$-th failure injection into the $j$-th deployment.
Although the full recorded dataset can be replayed, to reduce resource utilization during profiling, the user can specify a time interval where events just prior to and after the points of failure injection can be replayed thus limiting time spent in this phase.
At the same time, performance metrics are gathered across all parallel deployments in order to measure recovery times. 
The metrics we are most concerned with include: 

\begin{itemize}[leftmargin=*]
    \setlength{\itemsep}{3pt}
    
    \item{\verb|Input Throughput|}: Measured in events per second, this value represents the sum of the events entering the source operators of the DSP job per second.

    \item{\verb|Average Consumer Lag|}: Measured in number of events, this value represents the number of events accumulated at the messaging queue waiting to be consumed by the source operators of the DSP job.
    
\end{itemize}

In order to measure how long it takes for the DSP job to recover after experiencing a failure, the aforementioned metrics are used to train an anomaly detection algorithm on positive executions, i.e. let the function $s : X \rightarrow X$ perfectly represent the metrics data stream such that for any given data point $x \in X$ the prediction is always $s(x) = x$. 
Assuming most data collected in the recording period is normal, the algorithm can learn to detect deviations from the expected normal behaviors and therefore will report an anomalous behavior if a configurable threshold based on a window of past errors is hit. 
Measuring the length of time the system was in an anomalous state is therefore equivalent to the recovery time. 
In order to accomplish this, we utilize an online ARIMA method motivated in~\cite{SSG+18} for our implementation. 

It is important to note that in this context, the recovery time is not only referring to the time the system is in an inconsistent state before processing resumes. 
For systems employing checkpoint and rollback recovery strategies, processing will start at a previous saved offset and then attempt to catch up to the latest offset, all while new events are continuously arriving. It is this length of time, from when the failure occurs to the point at which processing is once again producing results at the latest offset, that we are interested in measuring, as it is a more accurate reflection of availability.
Thus, we define the set of recovery times as measured by the anomaly detector as

\begin{equation}
    R = \{r_i^{(j)}| 1 \leq i \leq m, 1 \leq j \leq z\}.
\end{equation}

Consequently, the lengths of the observed recovery times in $R$ are influenced by two main factors: the variable nature of the workload, i.e. the changing number of messages arriving per second over time; and the point at which the failure occurs relative to the next checkpoint completed successfully. 
As we intend to create a prediction model in the next phase using the recovery time observations, these factors need to be considered as they introduce variance influencing comparability, thus making our models unusable without further adaptions. 
Concerning the workload, we make the assumption that over these shorter time periods where recovery takes place, i.e. less than 15 minutes, the changing throughput rates would average out over time. Therefore, a constant workload can be assumed
However, the point at which the last checkpoint completed successfully will differ across failures and will directly impact the number of messages that need to be reprocessed, thus increasing or decreasing the recovery time. Therefore, for the purposes of our work, we only consider the worst-case scenario, i.e. we assume failures occur at the point right before the next checkpoint completes successfully. As such, for profiling runs, we measure the distance in time until the next checkpoint is scheduled to start, and inject the failure just prior to this point. 





At the conclusion of the profiling runs, the parallel deployments are deleted and the resources are released, with the profiling sets $C$, $TR$, $L$, and $R$ being passed into the third and final phase addressing modeling and runtime optimization.

\subsection{Phase 3: Modeling and Runtime Optimization}

The third and final phase is intended to execute indefinitely while continuously optimizing the targeted DSP job by monitoring for violations of two user-defined QoS constraints: $l_{\text{const}}$ which defines an upper bound on the average end-to-end latency; and $r_{\text{const}}$ which defines an upper bound on predicted recovery time. 
A violation of either constraint triggers a reconfiguration of the system where a new CI is chosen.
Care must be taken when choosing a new CI value, as increasing the frequency with which checkpoints are performed will result in better recovery times but could likewise negatively impact latencies and vice versa.
Therefore, we formulate this as an optimization problem where the objective is to select a CI that minimizes for both performance and availability.
A graphical representation can be seen in \autoref{fig:overview}(c).

In order to achieve this, we train two multiple regression models using the data gathered in the preceding two phases.
The performance model $M_L$ aims at finding a mapping such that $M_L: C,TR \rightarrow L$, whereas the recovery time model $M_{R}$ is configured as $M_{R}: C,TR \rightarrow R$. 
Once both models have been trained with our observed values, metrics from the targeted DSP job are continuously gathered and evaluated. 
For performance violations, Khaos compares the current average end-to-end latency to the performance constraint $l_{\text{const}}$.
As end-to-end latencies tend to be quite volatile, our previously fitted models likely contain noise for which we need to account when making actual predictions. 
Thus, we employ a correction approach for the prospective prediction error to localize predictions to the current cluster conditions.
Khaos keeps track of the latency observations over the past $k$ optimization iterations, averages across the $k$ pairwise fractional differences given the current latency, and then uses this estimated rescaling factor $p$ to rescale the predicted value obtained from our model.     
For recovery time violations, it determines the average throughput rate and together with the current CI uses $M_{R}$ to predict the recovery time considering the worst case scenario.

However, reconfiguration is not without its own cost and requires a full restart of the job.
Therefore, when violations of the constraints are detected, Khaos will determine whether or not a reconfiguration should be performed immediately or if this decision should be deferred until the next optimization cycle.
In order to achieve this, a TSF model is trained on the incoming message rate and a multi-step ahead forecast is performed should a violation be detected.
If the prospective incoming message rate is expected to decrease significantly, i.e. more than 10\%, from the current point in time until the next optimization run is executed, then the decision to reconfigure can be delayed. 
Otherwise, the reconfiguration can proceed as per normal.
The goal of reconfiguration is to select a CI value that results in the furthest distance from the two upper bounds that satisfies both.
Formally, this multi-objective optimization problem can be formulated as

\begin{equation}
    \begin{aligned}
    & \underset{C}{\text{min}} & & Q_R + Q_L^* + |Q_R - Q_L^*|\\
    & \text{s.t.} & & Q_R < r_{\text{const}},\\
    & & & Q_L^* < l_{\text{const}},\\
    & & & Q_R, Q_L^* > 0.
    \end{aligned}
\end{equation}
Here, $Q_R$ is the fraction $\frac{M_R(C, TR_{\text{avg}})}{r_{\text{const}}}$ between the prediction of the recovery time model and the corresponding constraint. 
The same applies to the latency model, except that $Q_L^*$ describes the fraction after rescaling, i.e. $Q_L^* = p \cdot Q_L$.
Given our configuration set $C$ and the current average throughput rate $TR_{\text{avg}}$, we aim at finding a value for the CI that satisfies both objectives individually.
If minimizing the above expression finds a new value for the CI, which is predicted to be more performant, then the system is reconfigured should the expected workload not decrease and monitoring continues.

%% file: sections/04_evaluation.tex
\section{Evaluation}

\begin{figure*}
\centering
\subfloat[IoT Vehicles Experiment Dataset.]{
  \includegraphics[width=\columnwidth]{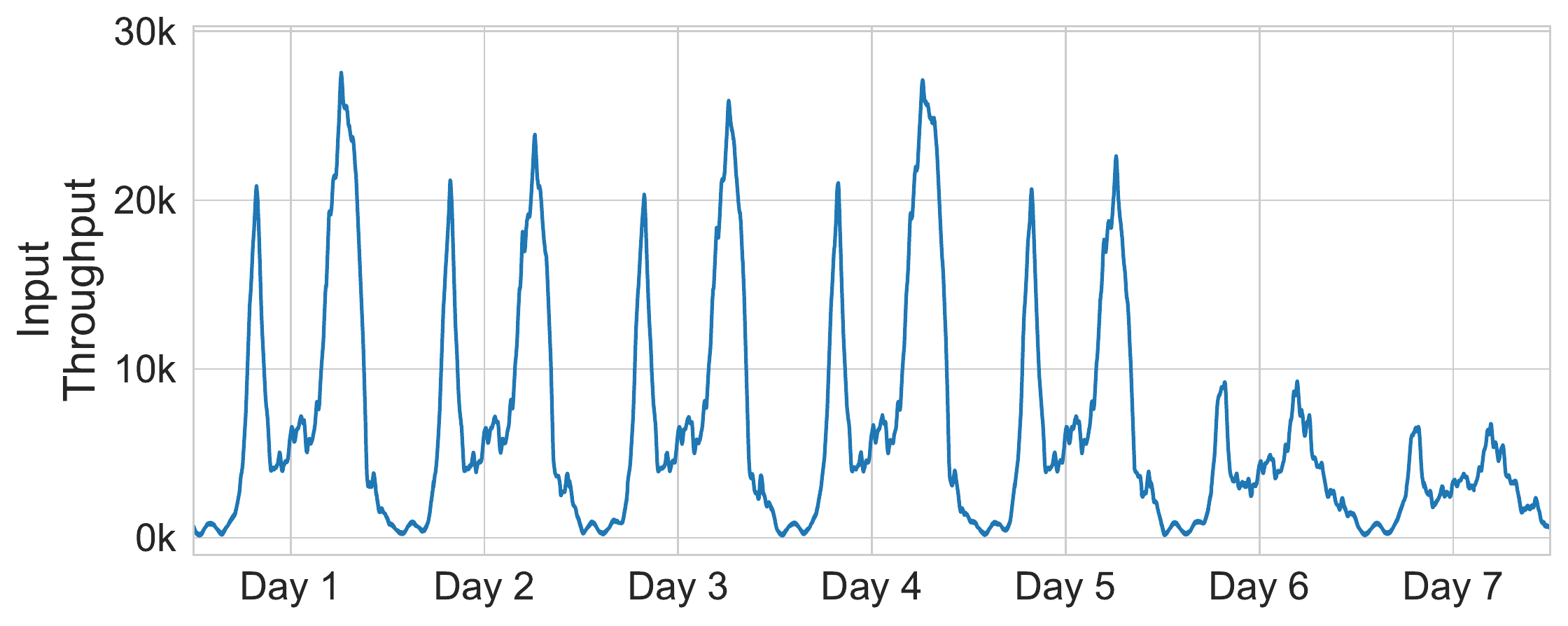}
}
\subfloat[YSB Experiment Dataset.]{
  \includegraphics[width=\columnwidth]{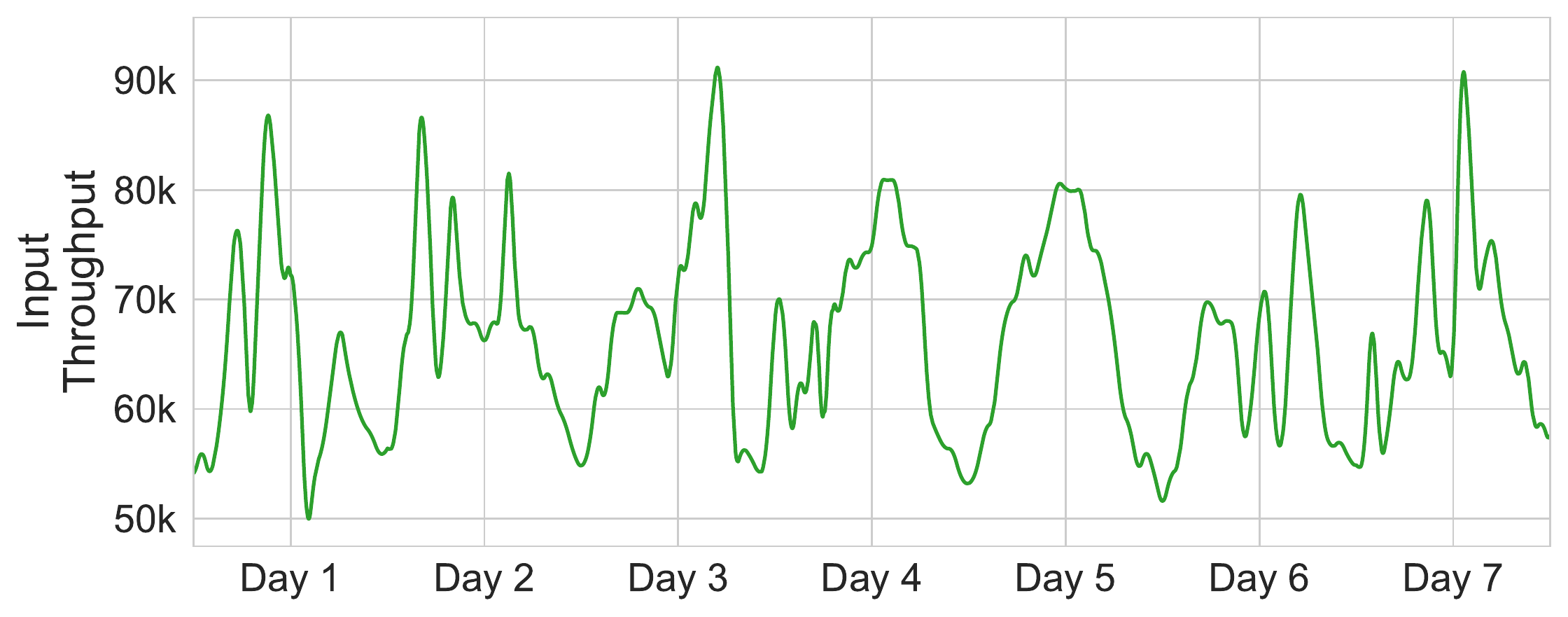}
}
\newline
\subfloat[IoT Vehicles Experiment. Failure injections and reconfigurations.]{
  \includegraphics[width=\columnwidth]{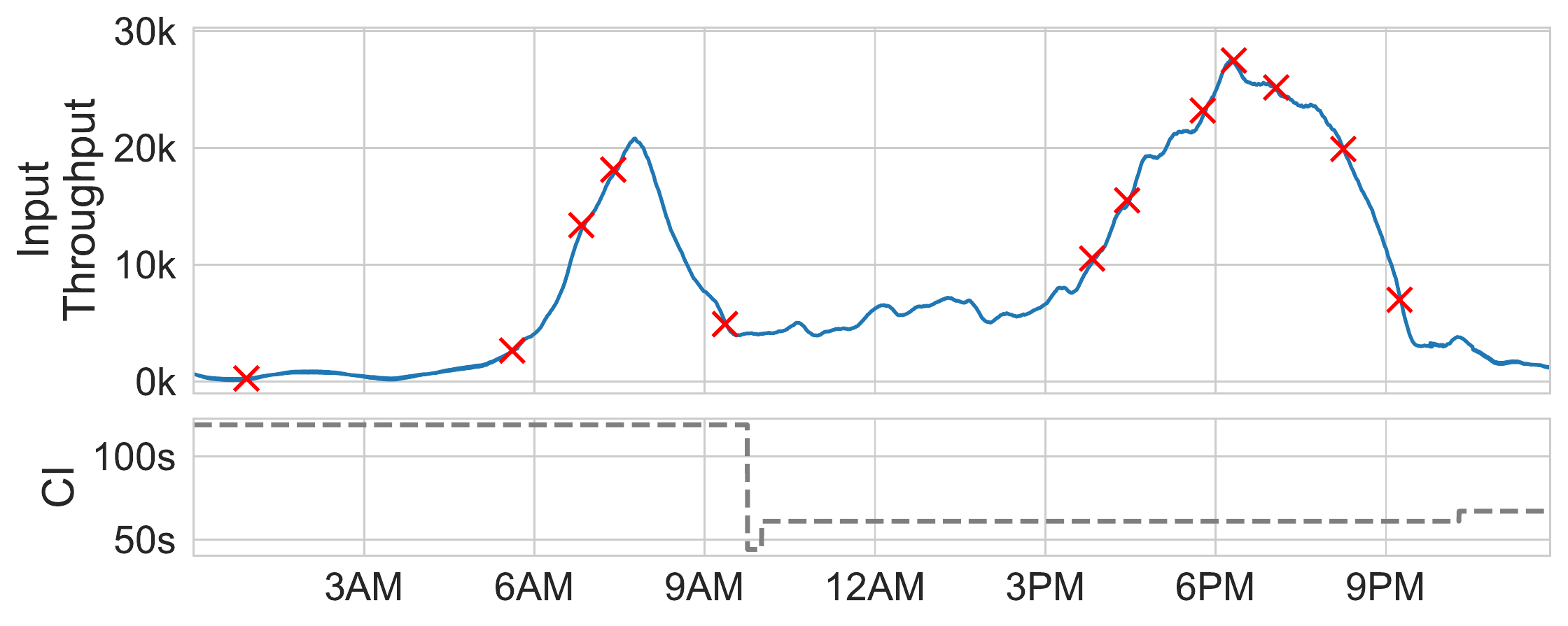}
}
\subfloat[YSB Experiment. Failure injections and reconfigurations.]{
  \includegraphics[width=\columnwidth]{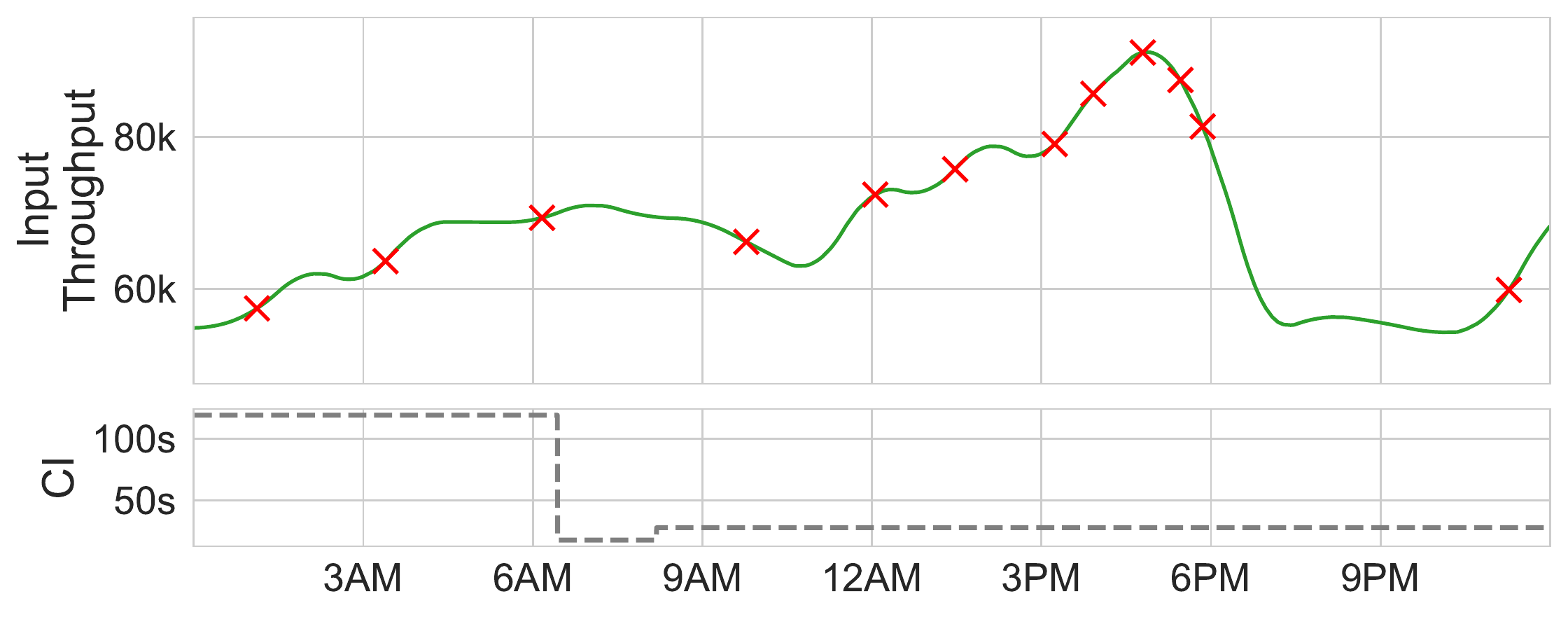}
}
\caption{Overview of datasets, their throughput rates, utilized failure injection points, and CI reconfigurations triggered by Khaos.}
\label{fig:evaluation_datasets}
\end{figure*}

Now we show that Khaos is both practical and beneficial for DSP through experimentation. The prototype, data, and experiment artifacts can be found in the following repository\footnote{\url{https://github.com/dos-group/khaos}}.

\subsection{Experimental Setup}

\begin{table}[ht]
\centering
    \begin{tabular}[t]{rp{0.65\linewidth}}
        \toprule
        Resource&Details\\
        \midrule
        OS&Ubuntu 18.04.3\\
        CPU&Quadcore Intel Xeon CPU E3-1230 V2 3.30GHz\\
        Memory&16 GB RAM\\
        Storage&3TB RAID0 (3x1TB disks, linux software RAID)\\
        Network&1 GBit Ethernet NIC\\
        Software&Java v1.11, Flink v1.12, Kafka v2.6, ZooKeeper v3.6, Docker v19.3, Kubernetes v1.18, HDFS v2.8, Redis v5.0, Prometheus v2.25
        \\
        \bottomrule
    \end{tabular}
\caption{Cluster Specifications}
\label{tbl:clusterspecs}
\end{table}%

Our experimental setup consisted of a co-located 50-node Kubernetes~\cite{VPK+15} and HDFS~\cite{SKR+10} cluster as well as a 3-node Apache Kafka\footnote{\url{https://kafka.apache.org/}, Accessed: May 2022} cluster configured to have 8 partitions and a replication factor of 3. 
Node specifications and software versions are summarized in~\autoref{tbl:clusterspecs}. A single switch connected all nodes. 
Each experiment consisted of a Kubernetes namespace containing: an Apache Flink\footnote{\url{https://flink.apache.org/}, Accessed: May 2022} native session cluster with all jobs set to have a parallelism of 4; and a single Prometheus\footnote{ \url{https://prometheus.io}, Accessed: May 2022} time series database was used for the gathering of metrics. 
The Yahoo Streaming Benchmark (YSB) experiment additionally made use of a Redis\footnote{\url{https://redis.io/}, Accessed: May 2022} database. 
Regarding end-to-end latency measurements, averages were taken over the $99^{th}$ percentile in order to filter outliers during normal failure free operations. 
The timeout interval for Flink taskmanager nodes is 50s as per the default settings.
Importantly, for both experiments QoS constraints for performance and availability were set at 1000ms for end-to-end latencies and 240s for recovery times respectively.
Each experiment was conducted 5 times with the median selected for our results and discussion.

\subsection{IoT Vehicles Experiment}

We created a simulation that mapped the streets and intersections of an area of central Berlin, Germany. 
In this area a number of vehicles were generated travelling along various routes and providing an update message every 1 second. 
Each update message contained the vehicle\_ID, vehicle\_type, geolocation, speed, direction, and event\_time. 
This IoT Vehicles streaming dataset was generated using Sumo~\cite{SUMO2018} and the number of concurrent vehicles, i.e. the workload, is based on the TAPASCologne scenario\footnote{\url{https://sumo.dlr.de/docs/Data/Scenarios/TAPASCologne.html}; Accessed: May 2022}.
For this experiment, a 7-day streaming dataset was generated using random seeds to create variability across the various days.
Throughput rates over time can be seen in \autoref{fig:evaluation_datasets}(a).

A DSP job was created that processes the streaming vehicle data. 
It consisted of the following streaming operations: read an event from Kafka; deserialize the JSON string; filter update events not within a certain radius of a designated geo-point where vehicles are to be monitored; take a 3s sliding window with a slide of 1s where all update events are of the same vehicle ID and calculate the vehicle's average speed; generate a notification for vehicles that have exceeded the speed limit; enrich notification with vehicle type information from data stored in system memory and write it back out to Kafka.

\begin{table}
\caption{IoT Vehicles Experiment Results.}
\label{tbl:iotresults}
\centering
    \subfloat[Error Analysis.] {
    \begin{tabular}{p{23mm}|p{17mm}p{17mm}} \toprule
        {} & {$Performance$} & {$Availability$} \\ \midrule
        {Avg. Percent Error} & 0.099 & 0.131 \\ \bottomrule
    \end{tabular}
    }
    \newline
    \subfloat[IoT Vehicles Experiment Results.] {
    \begin{tabular}{p{21mm}|p{5mm}p{5mm}p{5mm}p{5mm}p{5mm}p{5mm}} \toprule
        {Configuration} & {Khaos} & {10s} & {30s} & {60s} & {90s} & {120s} \\ \midrule
        {Avg. Latency (ms)} & 737 & 1086 & 729 & 796 & 697 & 692 \\ \midrule
        {Lat Violations (\%)} & 0.087 & 0.153 & 0.062 & 0.110 & 0.073 & 0.060 \\ \midrule
        {Recovery Time (s)} & 2071 & 2757 & 1681 & 2064 & 2505 & 2904 \\ \midrule
        {Rec Violations (s)} & 197 & 1188 & 147 & 227 & 555 & 826 \\ \bottomrule
    \end{tabular}
    }
\end{table}

\subsection{YSB Experiment}

This experiment is based on the Yahoo Streaming Benchmark\footnote{\url{ https://yahooeng.tumblr.com/post/135321837876/benchmarking-streaming-computation-engines-at}, Accessed: May 2022}. It implements a simple streaming advertisement job where there are a number of advertising campaigns and a number of advertisements for each campaign.
The authors of the benchmark created a Kafka Producer application that would generate a constant stream of events containing, among other things, an event\_time, an event\_type, and an ad\_id. 
We created a generator that was combined with a click-through rate dataset\footnote{\url{https://www.kaggle.com/c/avazu-ctr-prediction}, Accessed: May 2022} to create the workload for this experiment.
This workload can be in~\autoref{fig:evaluation_datasets}(b).

The job of the benchmark is to read various JSON events from Kafka, identify relevant events, and store a windowed count of these events per campaign in Redis. The job consists of the following operations: read an event from Kafka; deserialize the JSON string; filter out irrelevant events (based on type field), take a projection of the relevant fields (ad\_id and event\_time),  join each event by ad\_id with its associated campaign\_id stored in Redis; take a 10s windowed count of events per campaign and store each window in Redis along with a timestamp of when the window was last updated. For the purposes of our experiments, we modified the Flink benchmark by enabling checkpointing and replacing the handwritten windowing functionality with the default Flink implementation. Although doing so decreases the update frequency to the length of each window, results should be accurate and more interesting for our experiments due to the accumulated windowing operator state at each node. 

\begin{table}
\caption{YSB Experiment Results.}
\label{tbl:ysbresults}
\centering
    \subfloat[Error Analysis.] {
    \begin{tabular}{p{23mm}|p{17mm}p{17mm}} \toprule
        {} & {$Performance$} & {$Availability$} \\ \midrule
        {Avg. Percent Error} & 0.122 & 0.0728 \\ \bottomrule
    \end{tabular}
    }
    \newline
    \subfloat[YSB Experiment Results.] {
    \begin{tabular}{p{21mm}|p{5mm}p{5mm}p{5mm}p{5mm}p{5mm}p{5mm}} \toprule
        {Configuration} & {Khaos} & {10s} & {30s} & {60s} & {90s} & {120s} \\ \midrule
        {Avg. Latency (ms)} & 653 & 691 & 660 & 637 & 576 & 527 \\ \midrule
        {Lat Violations (\%)} & 0.059 & 0.061 & 0.069 & 0.058 & 0.033 & 0.024 \\ \midrule
        {Recovery Time (s)} & 2319 & 2182 & 2126 & 2548 & 3093 & 3532 \\ \midrule
        {Rec Violations (s)} & 9 & 117 & 32 & 77 & 401 & 764 \\ \bottomrule
    \end{tabular}
    }
\end{table}

\begin{figure}
    \centering
    \subfloat[IoT Vehicles Experiment.]{
      \includegraphics[width=\columnwidth]{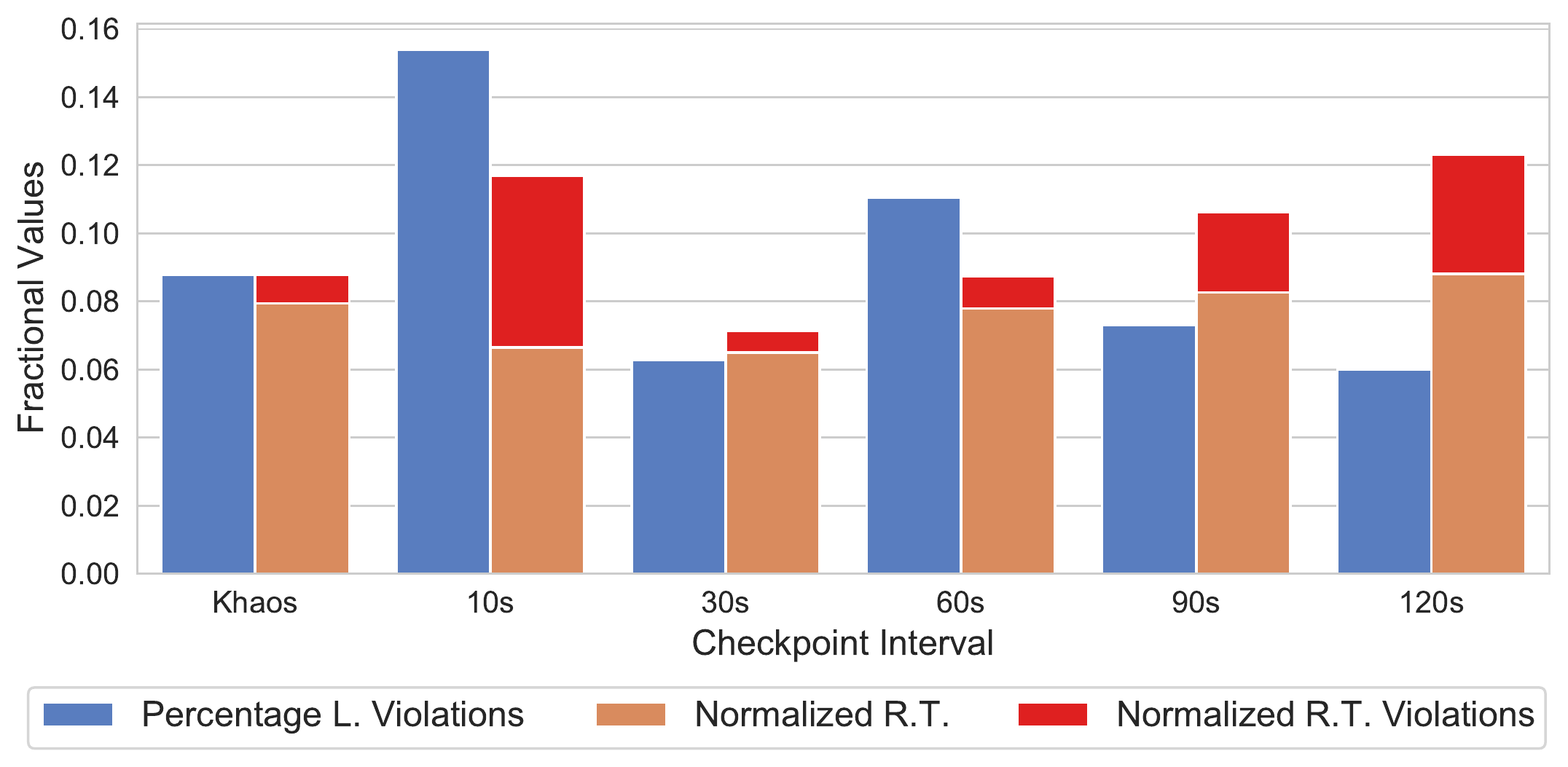}
    }
    \newpage
    \subfloat[YSB Experiment.]{
      \includegraphics[width=\columnwidth]{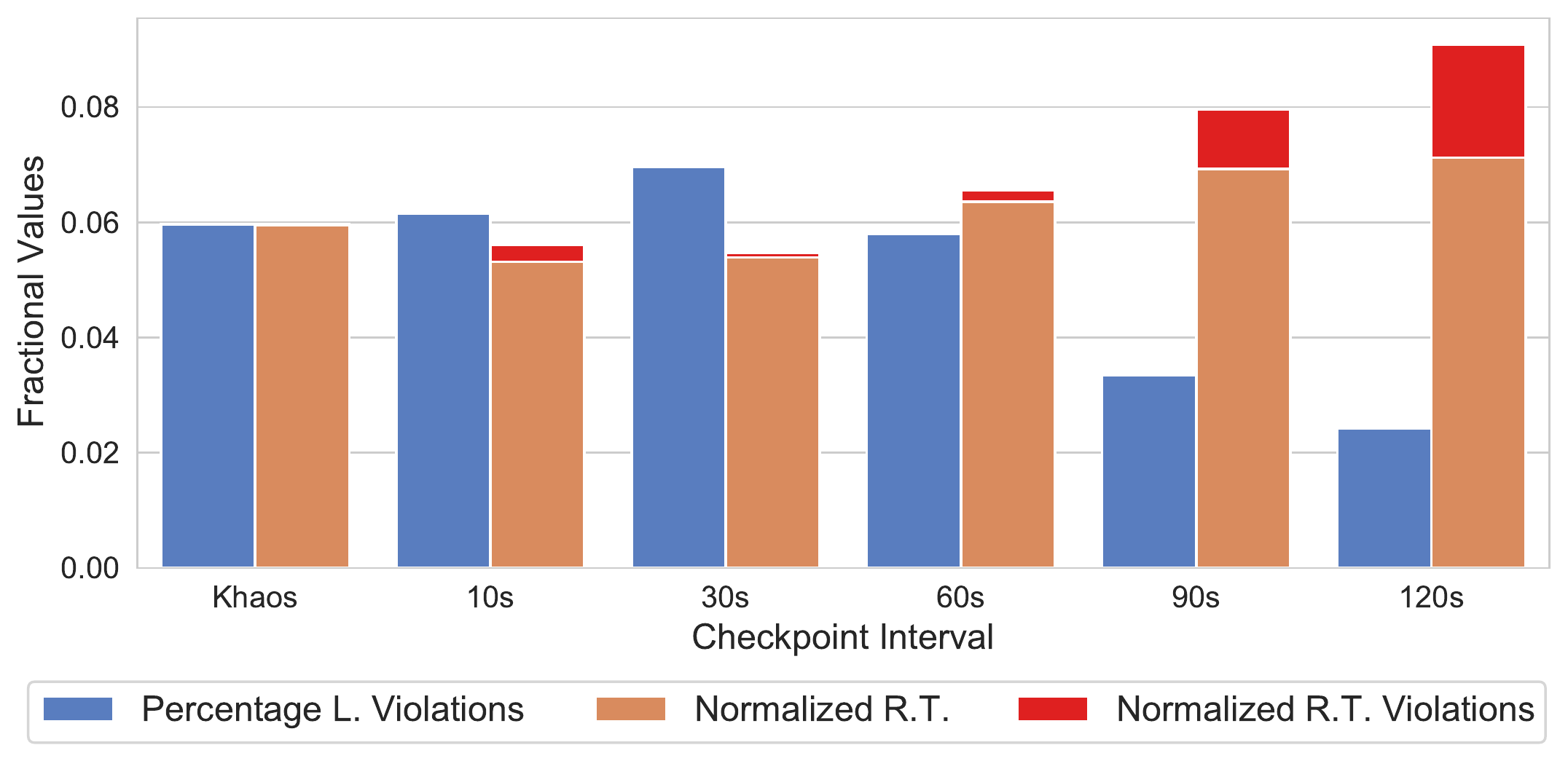}
    }
    \caption{Latency (L.) violations, normalized recovery times (R.T.), and normalized recovery time violations.}
    \label{fig:evaluation_violations}
\end{figure}

\subsection{Experimental Results \& Discussion}

The results of both experiments will now be presented and discussed in further detail.
The effectiveness of our approach was evaluated against 5 baseline runs, which were executed in parallel to ensure cluster conditions were commensurate.
For baseline runs, a range of static CI values were selected and streaming jobs started using these configurations.
Each job consumed the same stream of events.
Static CI values included 10, 30, 60, 90, and 120 seconds respectively.
Over the course of each experiment, 12 failures were injected at similar times across all jobs. 
Failure selection was based on different throughput levels which can be seen for both experiments in \autoref{fig:evaluation_datasets}(c) and \autoref{fig:evaluation_datasets}(d).
To ensure results were comparable, failures were injected at the end of the CI.
This is inline with our assumption of the worst case scenario, i.e. a failure occurs just before the next checkpoint completes successfully and therefore maximizing recovery times.

Concerning the predictive models trained as part of the profiling phase for each experiment, the results of a post execution error analysis can be seen in \autoref{tbl:iotresults}(a) and \autoref{tbl:ysbresults}(a).
These results measured the average percent error between predicted and actual observations for both latencies and recovery times across the full duration of the experiments.
Latency values were collected as part of the optimization loop and recovery times were measured when failures were injected.
The results show that on average latency predictions were within 9\% and 12\% of expected and recovery time predictions were within 13\% and 7\% of expected.
Considering the presence of this error within our predictions, we conclude that it was accurate enough for the optimization step to make good CI selections. 
We previously described our failure injection procedure highlighted in~\autoref{fig:evaluation_datasets}(c) and~\autoref{fig:evaluation_datasets}(d), we will now describe the bottom section of this figure. 
Associated results for these experiments can be seen in \autoref{tbl:iotresults}(b) and \autoref{tbl:ysbresults}(b).
During optimization, Khaos triggers reconfigurations of the CI in order to account for changes in latency and prospective recovery times.
We observed in total three reconfigurations for the IoT Vehicles Experiment, and two reconfigurations for the YSB Experiment.
It can be seen that the CI is often set to a lower value with increasing throughput rates. 
This is expected, as formulated recovery time constraints can likely not be fulfilled when maintaining a high CI value. 
In general, the few reconfigurations result from our method requiring one of the constraints to be fulfilled in order to conduct an optimization step. Consequently, reconfigurations are applied sparsely, and CI configurations in-between observed violations are not guaranteed to be optimized if both constraints are intermittently jointly violated.
An example for this can be seen in~\autoref{fig:evaluation_violations}(a) with recovery time violations for Khaos, indicating that CI updates were aborted at some point.

Regarding the violation of formulated performance and availability constraints, we report in~\autoref{fig:evaluation_violations} the fraction of time the latency constraint was not fulfilled, and normalize the measured recovery times accordingly such that the resulting bars are aligned for Khaos. This allows us to better assess the relation of both objectives for the utilized static CI values, i.e. our baselines.  
It can be clearly showed that the CI has an impact on latencies, i.e. with more frequent checkpoints being performed, latencies are higher and vice versa with recovery times.
This often leads to violations of the respective formulated constraints.
In our conducted experiments, this is less problematic for smaller CI values, which is also what Khaos often defaults to (see~\autoref{fig:evaluation_datasets}).
An exception to this is the 10s CI illustrated in~\autoref{fig:evaluation_violations}(a), which performs poorly due to a failure injection during catch-up from a previous failure. 
However, while certain static CI configurations show comparably good results, this is restricted to our exemplary streaming jobs only, i.e. the discovered CIs are not a general solution.
Based on the results presented in \autoref{tbl:iotresults}(b) and \autoref{tbl:ysbresults}(b), we can surmise that Khaos produces better average latencies overall than the high frequency CI configurations. Likewise the percentage of latency violations were commensurate with these configurations.
At the same time, Khaos produced fewer recovery time violations indicating that it provides a balance of both within the user-defined constraints.
It is important to note that while individual static configurations might marginally outperform Khaos for a specific job and workload, the benefits of such a selection will not generalize if the jobs and/or workloads change. 

%% file: sections/06_conclusion.tex
\section{Conclusion}

In this paper we presented Khaos, an approach which borrows from the principles of Chaos Engineering to allow for the automatic runtime optimization of DSP fault tolerance configurations.
It makes use of parallel profiling runs and failure injection to capture metrics, which are in turn used to model the performance and availability of targeted DSP jobs executing on variable workloads.
It does this in order to monitor for runtime violations of user-defined QoS constraints and, should a violation be detected, can search for and select near-optimal CI configurations, which provide a balance of both.
Through our experiments we showed that Khaos is able to optimize the CI configuration variable in order to minimize both latency and recovery time violations while outperforming most static configurations.
For future work we intend to investigate the feasibility of a continuous optimization routine which takes periods of low utilization into consideration.